\documentstyle[epsf]{cupconf}

\ifoldfss
\else
  \ifnfssone
    \newmathalphabet{\mathit}
      \addtoversion{normal}{\mathit}{cmr}{m}{it}
      \addtoversion{bold}{\mathit}{cmr}{bx}{it}
    \newmathalphabet{\mathcal}
      \addtoversion{normal}{\mathcal}{cmsy}{m}{n}
    \else
    \ifnfsstwo
    \fi
  \fi
\fi

\def\hexnumber#1{\ifcase#1 0\or1\or2\or3\or4\or5\or6\or7\or8\or9\or
 A\or B\or C\or D\or E\or F\fi }

\makeatletter
\ifx\CUP@mtlplain@loaded\undefined
\else
\fi
\makeatother
 \makeatletter
 \ifx\CUP@mtlplain@loaded\undefined
   \font\tenbmi=cmmib10 at 10pt
   \font\sevenbmi=cmmib10 at 7pt
   \font\fivebmi=cmmib10 at 5pt

   \newfam\bmifam
   \textfont\bmifam=\tenbmi
   \scriptfont\bmifam=\sevenbmi
   \scriptscriptfont\bmifam=\fivebmi
   
 \fi
 \makeatother
\ifnfsstwo

\fi
\ifnfssone

\fi
\ifoldfss

\fi

\mathchardef\varLambda="0103

\makeatletter
\ifx\CUP@mtlplain@loaded\undefined
\else
\fi
\makeatother
\makeatletter
\ifx\CUP@mtlplain@loaded\undefined
  \font\tenbms=cmbsy10
  \font\sevenbms=cmbsy10 at 7pt
  \font\fivebms=cmbsy10 at 5pt
  \newfam\bmsfam
  \textfont\bmsfam=\tenbms
  \scriptfont\bmsfam=\sevenbms
  \scriptscriptfont\bmsfam=\fivebms

  \edef\bsy@{\hexnumber\bmsfam}
  \mathchardef\bnabla="0\bsy@72
\fi
\makeatother
\title[Violent relaxation in hierarchical clustering]{Violent
relaxation in hierarchical clustering}

\author[S. D. M. White]%
{S\ls I\ls M\ls O\ls N\ns D.\ns M.\ns W\ls H\ls I\ls T\ls E$^1$}

\affiliation{$^1$Max-Planck Institut f\" ur Astrophysik,
Karl-Schwarzschild-Strasse 1, 85740 Garching bei M\"unchen, Germany}

\begin{document}
\ifnfssone
\else
  \ifnfsstwo
  \else
    \ifoldfss
      \let\mathcal\cal
      \let\mathrm\rm
      \let\mathsf\sf
    \fi
  \fi
\fi

\maketitle

\begin{abstract}
The term ``violent relaxation'' was coined by Donald Lynden-Bell as 
a memorable oxymoron describing how a stellar dynamical system
relaxes from a chaotic initial state to a quasi-equilibrium. His 
analysis showed that this process is rapid, even for systems with 
many stars, and that it leads to equilibria which may plausibly be 
related to bounded isothermal spheres. I 
review how numerical simulations have improved our understanding of 
violent relaxation over the last thirty years. It is clear that the
process leads to equilibria which depend strongly on the initial
state, but which nevertheless have certain common features. A
particularly interesting case concerns objects formed in an
expanding universe through dissipationless hierarchical clustering 
from gaussian initial conditions; these may correspond to galaxy
clusters or to the dark halos of galaxies. While such objects display
a wide range of shapes and spins, the distributions of these
properties depend only weakly on the cosmological context and
on the initial spectrum of density fluctuations. Halo density 
profiles appear to have a universal form with a singular central
structure and a characteristic density which depends only on 
formation epoch. Low mass halos typically have earlier formation 
times and thus higher characteristic densities than high mass halos.
\end{abstract}

\firstsection 
\section{Introduction}

By the mid-1960's it was known that the luminosity
profiles of elliptical galaxies are smooth, symmetric, and well
fitted by theoretical models based on modifications of the isothermal
sphere (see, for example, King 1966). This was perceived as
paradoxical since Chandrasekhar's (1942) calculation of the effects
of ``collisional'' relaxation showed the relevant timescales for
elliptical galaxies to be much longer than the age of the universe.
Furthermore such two-body relaxation should lead to equipartition of energy
and so to radial segregation of stars by mass; this would produce
unacceptably large colour gradients in an elliptical galaxy.
Thus a relaxation process with a short timescale and no dependence 
on stellar mass is required to explain the observations.

Although Donald Lynden-Bell was not the first to realise that 
rapid changes in gravitational potential during protogalactic
collapse could provide such a process, his 1967 paper gave a 
clear description of the mechanisms involved, calculated the relevant 
timescale, showed how evolution might lead to a state resembling
an isothermal sphere but with no mass segregation, gave a first 
discussion of the reasons why real galaxies would be unable to reach 
the ``most probable'' state, and above all gave the process a name
that everyone can remember. This paper made strong claims in the
inimitable Lynden-Bell style, all of them stimulating, but not all, 
perhaps, fully justified. After deriving a fourth kind of equilibrium 
statistics to
set beside those of Maxwell-Boltzmann, Einstein-Bose, and Fermi-Dirac,
and arguing that incomplete relaxation produces rotationally
flattened objects resembling real galaxies, the text concludes by
throwing doubt on its own major assumptions and careering off into a 
discussion of the gravothermal catastrophe, the heat death of the 
universe, and the origin of the Seyfert phenomenon!

The forthright and provocative style of Lynden-Bell (1967) is
undoubtedly responsible both for its success and for its
rather mixed reputation among professional dynamicists. The latter can
be inferred from the slightly hostile (and in my opinion misguided) 
reworking in Shu (1978), and from the absence of any discussion
of the statistics of violent relaxation either in the standard textbook 
of Binney \& Tremaine (1987) or in Tremaine,
H\'enon \& Lynden-Bell's (1986) treatment of mixing processes during 
violent relaxation. There can be no doubt, however, that the 1967
paper did much to shape our current understanding of how stellar dynamical
systems come to equilibrium, and even today it remains both frequently cited 
and controversial. (See Earn's contribution to this volume for some 
illuminating statistics).

Numerical experiments in the 1970's showed that violent relaxation from
realistic initial conditions does {\it not} lead to a universal final 
structure, even after accounting for the restriction already discussed
explicitly by Lynden-Bell (1967), namely that the finite mass of any real 
system implies that the most probable statistical distribution can only be
realised for orbits with periods shorter than the collapse time of the
system, and so for energies more negative than some truncation threshold.
The discrepancy was first clearly demonstrated by the collapse simulations
of White (1976) and Aarseth \& Binney (1978). These showed that the violent
collapse of stellar dynamical systems can produce strongly aspherical
equilibria, whereas Lynden-Bell's arguments imply a final state which 
is spherical, at least in the absence of significant rotation. In numerical 
experiments the final state clearly remembers nonspherical aspects of the 
collapse and relaxation phases. Unfortunately the corresponding constraints
are difficult to formulate and have so far been ignored in attempts 
to go beyond the Lynden-Bell's original analysis and to obtain a deeper
understanding of the observed uniformity of elliptical galaxies ({\it e.g.} 
Stiavelli \& Bertin 1987; White \& Narayan 1987; Tremaine 1987). The only
clear area of progress since 1967 is in understanding the light profiles
at large radii. A number of authors have shown that asymptotically a relation 
$S\propto r_p^{-3}$ should hold between surface brightness $S$ and projected
radius $r_p$ (Aguilar \& White 1986; Jaffe 1987; Tremaine 1987); this results
from the expected continuity in the distribution of stellar binding energies
across the value corresponding to escape from the system. 

In the current contribution I will discuss recent work on violent
relaxation in a different context, that of the formation of dark
halos by hierarchical clustering of dissipationless dark matter
in an expanding universe. Most past studies of this process have
concentrated on differences in the predicted halo structure as a function of
the density of the Universe, and of the nature of
the initial fluctuations from which structure forms. Here I will instead
highlight certain regularities that emerge from the available 
numerical data, and argue that in this context violent relaxation
actually produces a ``universal'' density structure which is independent
of halo mass, of cosmological parameters, and of the initial fluctuation
spectrum. Unfortunately, we do not yet have a physical understanding of 
this universal structure of the kind which Lynden-Bell (1967) attempted 
to provide.

\section{Hierarchical clustering}

An idealised model for the growth of structure in the universe makes the
following assumptions.\hfil\break
\noindent{(i) The dominant mass component in the universe is the dark matter.
This consists of particles which interact only through gravity and whose
individual masses are so small that pairwise encounters can be neglected
when studying evolution on galactic and supergalactic scales. Dark matter
candidates which satisfy this assumption include massive neutrinos, axions, 
the lightest supersymmetric particle, ``jupiters'', and black holes of
masses 100$M_\odot$ or less. The distribution of dark matter can then be
described by its one-particle phase-space distribution $f(x,v, t)$
which evolves according to the collisionless Boltzmann equation, 
$Df/Dt=0$.}\hfil\break
\noindent{(ii) The evolution of the non-dominant component (which
we can actually see!) does not significantly affect that of the dark
matter distribution. This assumption clearly breaks down in the
inner regions of galaxies but may be acceptable on larger scales.}\hfil\break
\noindent{(iii) At early times the dark matter particles have a nearly
uniform spatial distribution and have small velocities relative to the
mean expansion of the Universe. The latter requirement eliminates light
neutrinos with masses $\sim 30$ eV since their velocities would not be
negligible on the relevant scales. The other candidates mentioned above
are still all acceptable.}\hfil\break
\noindent{(iv) If the dark matter density fluctuation field at early
times is represented as a superposition of plane waves, 
$\delta(x)=\int d^3k \delta_k \exp (ik.x)$, then 
the phases of different waves are
independently and randomly distributed on $(0,2\pi]$. 
The statistical properties of the random field $\delta( x)$ are then
fully represented by its power spectrum $P(k)\propto |\delta_k|^2$, 
and the distribution of $\delta(x)$ at any arbitrary set of 
positions $(x_1,x_2,....)$ is a multivariate gaussian. Such fields
are known as gaussian random fields and are predicted by a wide class of
theories for the origin of structure in the universe.}\hfil\break
\noindent{(v) The quantity $k^3P(k)$ increases with wavenumber $k$. This
is the condition that small objects form first and then aggregate into
larger systems as structure grows. All currently popular theories for 
structure formation satisfy this requirement on the scales relevant to 
galaxies and galaxy clusters.}\hfil\break

Over the last twenty years cosmological N-body simulations have been
used extensively to study nonlinear evolution from initial conditions 
obeying these assumptions. This kind of evolution is generically
referred to as ``hierarchical clustering''.  Here I ask whether
evolution from such initial conditions leads to structural regularities in
the highly nonlinear, quasi-equilibrium objects which it produces.
In concrete terms I ask whether dark halos formed by hierarchical 
clustering have a universal internal structure independent of their mass, 
of the density of the Universe, and of the initial power spectrum. 
Such universality could clearly be ascribed to the effects of violent 
relaxation as envisaged by Lynden-Bell (1967). Although cosmological
simulations have usually been used to compare predictions for
large-scale structure with the observed clustering of galaxies, a number 
of studies have also specifically investigated halo structure. The most 
useful references in this context are Efstathiou et al. (1988),
Frenk et al. (1988), Zurek et al. (1988), Dubinski \& Carlberg (1991), 
Warren et al. (1992), Crone et al. (1994), 
Navarro et al. (1995, 1996) and Cole \& Lacey (1996). An important additional 
reference is the analytic work of Hoffman \& Shaham (1985) and Hoffman 
(1988) which suggests that the slope of the density profile of a dark halo 
in its outer regions should depend systematically on the power spectrum 
of initial density fluctuations and on the density of the universe.

\section{Shapes and spins of halos}

Since the very earliest simulations it has been clear that hierarchical
clustering does not produce near-spherical halos. Rather the equidensity
surfaces of many simulated halos can be well fit by a set of concentric
ellipsoids with axial ratios differing substantially from unity. Major 
deviations from this structure can usually be ascribed to non-equilibrium
effects such as recent mergers. Halos can be oblate, prolate, or any 
intermediate shape, and their axial ratios can vary substantially with
radius. In the published literature there is some disagreement about
the details both of these variations and of the distribution of axis ratios.
It seems likely that this reflects, at least in part, the different
definitions and algorithms used in computing axial ratios. In fact,
a comparison of the results of Frenk et al. (1988), of Dubinski \& 
Carlberg (1991), of Warren et al. (1992) and of Cole \& Lacey (1996) 
shows that the distribution of halo shapes is similar in all cases. 
Very few objects are nearly spherical, and roughly half have ratios of
longest to shortest axis exceeding two. The intermediate axis length 
scatters rather uniformly beween the other two with a weak preference 
for near-prolate halos. Although some systematic differences are found
as a function of halo mass and of the initial power spectrum, these
trends are quite weak. 

The angular momentum distribution for dark halos was one of the
first of their properties to be systematically studied by numerical
simulation (Efstathiou \& Jones 1979). Almost all later work has
followed this original paper and has characterised the rotation rate
of a dark halo by its spin parameter, $\lambda=J|E|^{1/2}/GM^{5/2}$
where $J$, $E$ and $M$ are respectively the total angular momentum, 
energy and mass (Barnes \& Efstathiou 1987, Efstathiou et al. 1988, 
Warren et al. 1992, Cole \& Lacey 1996). There is general agreement that
the distribution of $\lambda$ is very broad with a median between
0.04 and 0.06 and a range of about a factor of 2.5 between the
20\% and 80\% points. The median declines very weakly with increasing 
halo mass and is about 10\% of the value expected for a fully
rotationally supported system such as a self-gravitating disk.
Within a dark halo the rotation velocity varies little with
radius and there is a relatively poor correlation
between the rotation axes at different radii. This is strongly at
variance with the solid body rotation predicted by Lynden-Bell (1967),
although Donald was careful to point out that his prediction could 
not apply far from the centre of a relaxed system. There is at most 
a very weak dependence of the $\lambda$-distribution on $P(k)$ or on 
$\Omega$, with different authors finding slightly different results. 
These differences probably reflect different (arbitrary) definitions 
for the boundary of a dark halo. 

Violent relaxation during hierarchical clustering clearly does not 
produce objects with a universal shape or spin. It is, however, quite
striking that the {\it distributions} of these quantities do seem to
be (almost) universal. Study of the formation of individual objects
suggests that both their shape and their spin depend on the details of
their nonlinear collapse and on their immediate environment at that
time. For example, the objects with the largest spin tend to be those
which formed by merging of two progenitors of comparable mass, while
extreme prolate systems are often related to collapse along filaments.
The universality of the spin and shape distributions thus reflects
some kind of statistical universality for nonlinear formation paths in
hierarchical clustering. Quite different distributions of shapes and
spins can be found in models which depart fundamentally from the
hierarchical clustering paradigm ({\it e.g.} White \& Ostriker 1990).

\section{Density profiles}

While the near-universality of the spin and shape distributions of dark
halos has always been generally accepted, discussions of halo
density profiles have tended to focus on the differences 
expected in different cosmological contexts. This may stem from
two influential papers by Hoffman \& Shaham (1985) and Hoffman
(1988). These used a spherical infall model to study the nonlinear
evolution of structure around a peak of an initially gaussian
overdensity field. The first paper showed that if the initial
fluctuation power spectrum is a power law, $P(k)\propto k^n$, and if
the universe is Einstein-de Sitter, then halos might be expected to
show power law density profiles, $\rho\propto r^{-\alpha}$ with
$\alpha=(9+3n)/(4+n)$ for $n\geq -1$ and $\alpha=-2$ for more negative
$n$. The second paper showed that steeper density profiles (larger effective
values of $\alpha$) are expected in universes with lower values of 
$\Omega$. Halo formation is, of course, very far from spherically 
symmetric in hierarchical clustering, so it was quite surprising
when a variety of simulation studies confirmed the predicted trends 
and even found reasonable quantitative agreement when $\alpha$ is 
estimated at radii where the density is a few hundred to a few 
thousand times the critical value (Quinn et al. 1986, Efstathiou et 
al. 1988, Zurek et al. 1988, Crone et al. 1994). None of this work
comments on any possible dependence of density profile on halo mass,
perhaps because the theoretical argument can easily be construed as
suggesting that no such dependence should be present.

Although the above studies all conclude that halo density profiles
reflect both $P(k)$ and $\Omega$ and so provide a possible route
to estimating these quantities, I will argue here that they
can also be considered a universal outcome of violent relaxation
during hierarchical clustering, and to be independent of the larger 
cosmological context in exactly the same way as the spin and shape
distributions. I will also argue that they have a substantial
dependence on halo mass which is at least as strong as the dependence
on $P(k)$ and $\Omega$.

In a high resolution study of the formation of rich clusters of
galaxies in the standard CDM cosmogony Navarro et al (1995) noticed that the
density profiles of the dark matter distribution in their simulated
clusters could be well described by a simple fitting formula:
\begin{equation}
{\rho(r)\over \rho_{crit}} = {\delta_c\over (r/r_s)(1+r/r_s)^2},
\end{equation}
where $\rho_{crit}$ is the critical density for closure, $r_s$
is a characteristic linear scale, and $\delta_c$ is a characteristic
overdensity. This profile is very similar to the analytic form which
Hernquist (1990) proposed as a convenient fit to the light
distribution in elliptical galaxies. Both formulae are singular for
small $r$ with $\rho\propto r^{-1}$; however equation (4.1) gives
$\rho\propto r^{-3}$ at large radii whereas the Hernquist law gives
$\rho\propto r^{-4}$. Note that Navarro et al apply their formula
only within the conventional ``virial radius'' $r_{200}$ 
of a cluster. This is defined to be the radius of the sphere within
which the mean density is 200 times the critical value, and it is a
good estimate of the region within which a cluster can be assumed to
be approximately in hydrostatic equilibrium (see Cole \& Lacey 1996).

In follow-up work Navarro et al (1996) found that equation (4.1) is
actually a good representation of the density structure of halos of
any mass in the standard CDM cosmogony, and that the characteristic 
overdensity is strongly correlated with halo mass; low mass halos have
systematically larger values of $\delta_c$ than high mass halos. I
illustrate this in Figure 1 which is taken from their paper. Rather
than showing the density profiles directly, this plot gives circular
velocity profiles for two halos. These are defined through
\begin{equation}
V_c(r) = (GM(r)/r)^{1/2},
\end{equation}
where $M(r)$ is the total mass contained in a sphere of radius $r$.
A characteristic mass and circular velocity for each halo can then be
defined through $M_{200}=M(r_{200})$ and $V_{200}=V_c(r_{200})$. In
Figure 1 the radius and the circular velocity are scaled to $r_{200}$
and $V_{200}$ respectively. The two halos shown have $M_{200}$ values
of $3\times 10^{15}M_\odot$ and $3\times 10^{11}M_\odot$ and so
correspond to a rich cluster and to a small galaxy halo. The
fact that the two scaled curves do not coincide shows that the density
profiles of the two objects differ by more than a simple linear
scaling. The small halo corresponds to the curve which rises highest
and so has a higher concentration and a larger characteristic
overdensity than the rich cluster. Both curves are well fit by
equation (4.1). The rich cluster curve could also be fit quite well by the
Hernquist formula, but it is clear from Figure 1 that this formula gives a
substantially poorer fit to the circular velocity curve of the small
halo. 
\begin{figure} 
  \epsfysize=27pc
  \epsffile{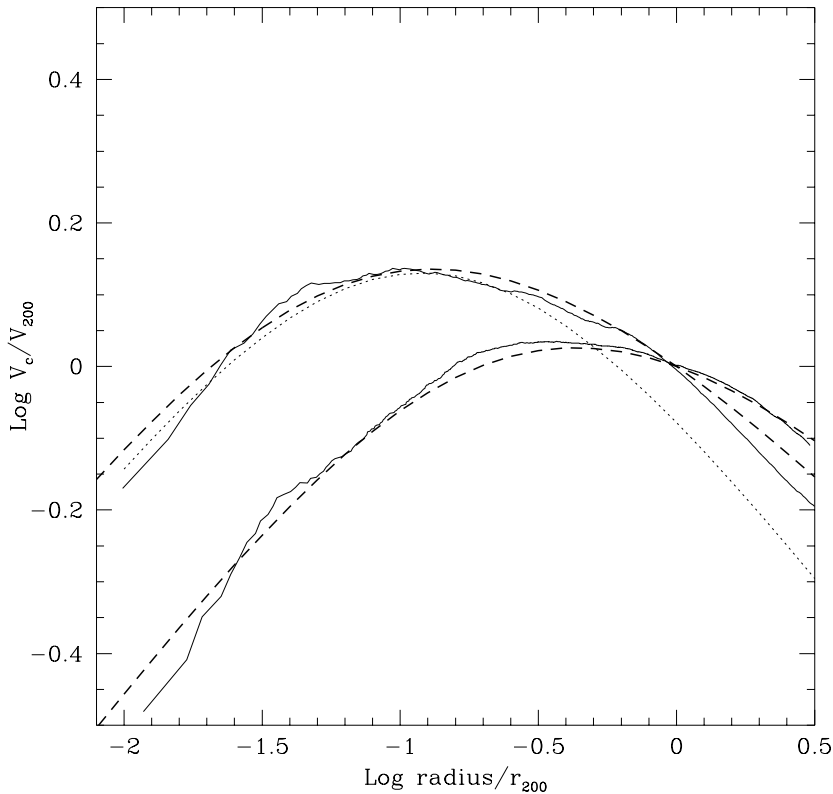}
  \caption{Scaled circular velocity curves for two simulated dark
halos in a standard CDM universe (solid curves). Velocities and radii
are normalised by $r_{200}$ and $V_{200}$ respectively, where
$(r_{200},V_{200})$ is (172kpc, 86 km/s) for the upper curve and
(3.7Mpc, 1860km/s) for the lower one. The dashed lines are the best
fits of equation (4.1) to the numerical data, while the dotted line is
the best fit of a Hernquist profile to the upper curve in the range
$r<r_{200}$.}
\end{figure} 

An obvious question is whether the good fit to equation (4.1) is
specific to CDM halos or reflects a more general property of
hierarchical clustering. In a beautiful recent paper Cole \& Lacey
(1996) show that this same fitting formula is a very good
representation of the mean shape of halo density profiles in
simulations of ``scale-free'' hierarchical clustering in universes 
with $\Omega=1$ and $P(k)\propto k^n$. (They consider $n$ values
in the range $-2\leq n\leq 0$.) In such universes the clustering 
patterns at different times are statistically equivalent once they 
are scaled in mass by the characteristic value $M_*(t)\propto 
t^{4/(3+n)}$. Cole \& Lacey demonstrate that for given $n$ the 
characteristic overdensity of dark halos decreases with increasing 
$M/M_*$, just as in the CDM case. Moreover they find that at fixed 
$M/M_*$ halo profiles have systematically smaller $\delta_c$ in 
universes with more negative $n$. It seems therefore that equation 
(4.1) is a good representation of the density profiles of dark halos for 
dissipationless hierarchical clustering in any Einstein-de Sitter universe.

Our own follow-up work (Navarro, Frenk \& White, in preparation) 
has confirmed Cole \& Lacey's results for halos
in scale-free universes and has extended the ranges both of halo mass and
of radius within a halo for which equation (4.1) is demonstrated to be a
good fit; at present these are roughly $0.01 M_*\leq M_{200}\leq 100M_*$ 
and $0.01r_{200}\leq r \leq r_{200}$. We have also tested our fitting
formula in low density universes (with and without a cosmological
constant) with both power law and CDM-like initial power spectra.
In all cases we have tried so far (which cover $0.1\leq \Omega\leq 1$)
the fit remains quite good. At given $M/M_*$ halos have higher 
overdensities in universes with smaller values of $\Omega$; at given
$M/M_*$ and $\Omega$ they have higher overdensities in
open universes than in flat universes with a cosmological constant.

\begin{figure} 
  \epsfysize=27pc
  \epsffile{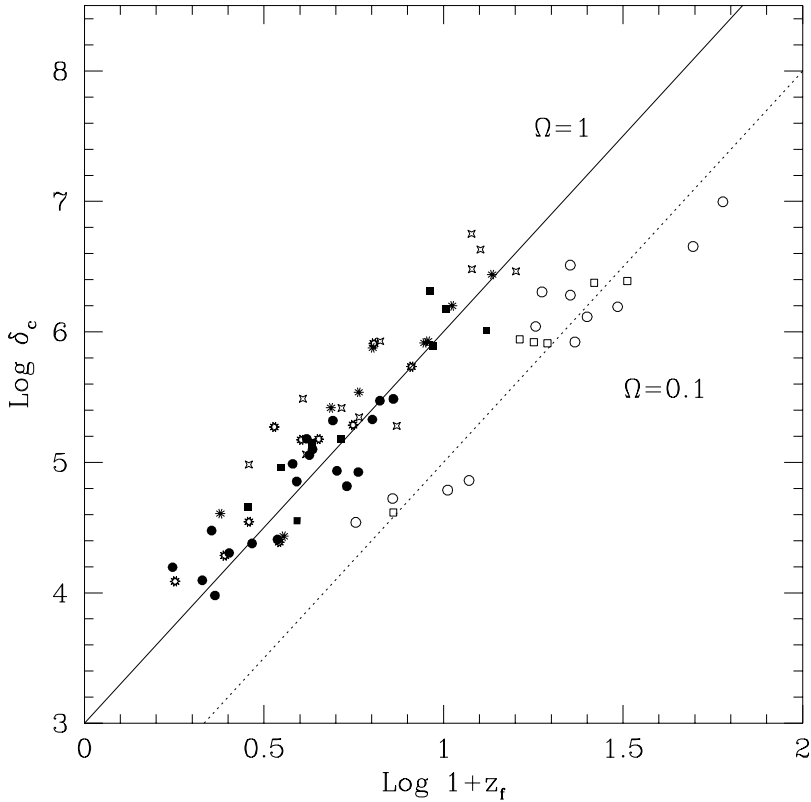}
  \caption{Characteristic overdensity is plotted against formation
redshift for simulated halos. Each symbol corresponds to a different
cosmogony as follows: open squares and open circles are halos in an
open universe, $\Omega_0=0.1$, with scale-free initial fluctuation
spectra with $n=0$ and $n=-1$ respectively; filled circles are halos
in a standard CDM cosmogony; other symbols refer to halos in
scale-free Einstein-de Sitter universes with $n=0, -0.5,
-1$ and $-1.5$. For each cosmogony halo mass decreases systematically
with increasing $\delta_c$ and $z_f$. The two straight lines
correspond to equation (4.3) for $\Omega=1$ and $\Omega=0.1$.}
\end{figure} 
 
It turns out that all these systematic dependences of halo concentration
on mass, on initial power spectrum, and on cosmological density can be
understood quite simply in terms of halo formation times. In hierarchical
clustering the high mass objects identified at any given time typically
formed much more recently than low mass objects identified at the
same time (see, for example, Lacey \& Cole 1993, 1994). Similarly, for a given
initial power spectrum $P(k)$ and nonlinear mass scale $M_*$,  halos of
some fixed $M/M_*$ form at higher redshift in a low density universe than
in an Einstein-de Sitter universe. Finally, in scale-free hierarchical
clustering  the growth of structure is faster (and so typical objects
form more recently) in universes with more negative values of the initial
power spectrum index $n$. With a suitable definition of formation redshift
$z_f$ all our data are consistent with the very simple relation,
\begin{equation}
\delta_c=1000\Omega_0(1+z_f)^3.
\end{equation}
This is shown in Figure 2 which was compiled by Julio Navarro using
all the data presently available from our programme of simulations. 
It is quite remarkable how well the value of $\delta_c$
scales with $(1+z_f)^3$ as measured directly in the simulations. (For
this plot  $z_f$ is defined for each halo as the earliest redshift at 
which its biggest progenitor has more than 10\% of its final mass. Other
definitions give similar results.) 

\section{Concluding remarks}

The material discussed in the preceding sections suggests that there is
indeed a certain universality in the properties of the objects which
form by dissipationless hierarchical clustering. Although it is
generally agreed that the shapes and spins of dark halos are predicted to
have distributions which depend at most weakly on halo mass, on the
power spectrum of initial density fluctuations, and on the density of
the universe, our claim that the density profiles of objects are also
independent of these quantities seems, at first sight, to conflict with
earlier work. This discrepancy is, however, only superficial.
Previous numerical studies fitted power laws to halo density profiles
over a limited range of overdensities, typically $\sim 10^2$ to $\sim 
10^4$. As a result, the systematic dependence of formation redshift on
power spectrum and $\Omega$, clearly visible from the point
distribution in Figure 2, produced systematic variations in the
measured slopes. Furthermore, most studies surveyed too small a range 
of halo masses to notice the systematic mass dependence. Direct 
comparison of our predictions with the numerical results of earlier 
studies shows good agreement in almost all cases.

It seems, therefore, that violent relaxation in 
hierarchical clustering leads to objects with universal density
profiles and universal distributions of shape and spin. Gravitational
potential fluctuations during the collision and merging processes
which characterize hierarchical clustering are evidently strong enough
to cause convergent evolution of the kind envisaged by
Lynden-Bell (1967). Since Donald's
detailed statistical arguments
provide at best a partial explanation of this process, further
work is needed to achieve a deeper understanding of the simplicity which
emerges from numerical experiment.

\begin{acknowledgments}
I would like to thank my collaborators, Julio Navarro and Carlos
Frenk, for permission to reproduce results from our joint research
programme. 
\end{acknowledgments}

\end{document}